\documentclass[12pt]{article}

\usepackage{epsfig}

\newcommand{\g}{\gamma}

\newcommand{\simgt}{\,\rlap{\lower 3.5 pt \hbox{$\mathchar \sim$}} \raise 1pt
 \hbox {$>$}\,}
\newcommand{\simlt}{\,\rlap{\lower 3.5 pt \hbox{$\mathchar \sim$}} \raise 1pt
 \hbox {$<$}\,}

{\end{list}}
\newcounter{enumct}

\begin{document}
 
\sloppy

\pagestyle{empty}
\begin{center}

{\LARGE\bf Virtual Photon Structure }\\[4mm]
{\LARGE\bf from Low $Q^2$ Jet Production\footnote{To appear in the 
    proceedings of the workshop on photon interactions and the photon
    structure, in Lund, September 1998}}\\[10mm]
{\Large G.~Kramer, B.~P\"otter} \\[3mm]
{\it II. Institut f\"ur Theoretische Physik,\footnote{Supported by
    Bundesministerium f\"ur Forschung und Technologie, Bonn, Germany, 
    under Contract 05~7~HH~92P~(0), and by EU Fourth Framework Program
    {\it Training and Mobility of Researchers} through Network {\it
      Quantum Chromodynamics and Deep Structure of Elementary
      Particles} under Contract FMRX--CT98--0194 (DG12 MIHT).}
 Universit\"at Hamburg,}\\[1mm]
{\it  Luruper Chaussee 149, D-22761 Hamburg, Germany}\\[1mm] 
{\it E-mail: kramer@mail.desy.de, poetter@mail.desy.de}\\[20mm]

{\bf Abstract}\\[1mm]
\begin{minipage}[t]{140mm}
We review next-to-leading order calculations of one-
and two-jet production in $ep$ collisions at HERA for photon
virtualities in the range $1< Q^2 < 100$ GeV$^2$. Soft and collinear
singularities are extracted using the phase space slicing
method. Numerical results are presented for HERA conditions with the
Snowmass jet definition. The transition between photoproduction and
deep-inelastic scattering is studied. We discuss the comparison with recent
H1 data of the dijet rate and differential dijet cross sections with special
attention to the region, in which two jets have equal transverse momenta. 
\end{minipage}\\[5mm]

\rule{160mm}{0.4mm}

\end{center}

\section{Introduction}

The interaction of the virtual photon with the constituents of the proton
in high energy $ep$ collisions is described differently depending how large
the scale of the photon virtuality $Q^2$ is. For large $Q^2$, i.e. 
large compared to other kinematical variables, as for example the
$E_T^2$ of produced jets, the virtual photon couples only directly to
the quarks originating from the proton. In the region
$0 \simlt Q^2 \simlt Q^2_{max}$ with $Q^2_{max}$ being small, the virtual
photon couples either directly to a parton from the proton
(so-called direct process) or through resolved processes, in which
the photon transforms into partons and one of these interacts with a
parton out of the proton to produce jets. On this basis the theory for
photoproduction ($Q^2 \simeq 0$) of jets at HERA and LEP, i.e. in $\g p$
and $\g \g $ processes, has been developped up to next-to-leading
order (NLO) QCD \cite{1} and good agreement with the experimental data from
HERA and LEP has been found \cite{2}. The cross section for jet production
is expressed as a convolution of universal parton distributions (PDF's) 
of the proton and, in the resolved case, of the
photon with the hard parton-parton scattering cross section. The evolution
of both parton densities with the scale $\mu$ as well as the hard
parton-parton scattering cross section can be calculated in perturbative
QCD as long as the scale $\mu $ of the hard subprocess, which is of the order
of the transverse energy $E_T$ of the produced jets, is large enough
as compared to $\Lambda_{QCD}$ and to $Q$. For these processes the
photon densities are defined for photon virtualities $Q^2 = 0$
and are constructed in such a way as to describe the wealth of data
in deep-inelastic $e\g  $ scattering or $\g ^* \g $ scattering,
where the photon $\g ^*$ has a large virtuality. 

For large $Q^2 \gg E_T^2$, i.e. in the deep inelastic region, the resolved
process is supposed to be absent and the hard electron-proton scattering 
cross section expanded in powers of $\alpha_s $ has to be convoluted only 
with the known PDF's of the proton. In this region jet production 
has been calculated with the help of several programs
\cite{3} and compared to data from the H1 and ZEUS collaborations \cite{4}
for various $Q^2$ regions. 

Thus we have a reasonably well tested theory for the photoproduction region,
i.e. regions of $Q^2$ which includes $Q^2 \simeq 0$, and the deep inelastic
region, where $Q^2$ is the large scale. Then the problem arises, how to
calculate jet production in $ep$ collisions with a fixed $Q^2 \neq 0$, 
although small compared to or at least less than the hard scattering scale
$\mu ^2$ ($Q^2 \simlt \mu^2$), which is usually taken to be of the order of
$E_T^2$. This challenging region of intermediate photon virtuality, where we 
have a typical two-scale problem, has been
considered by several authors in leading order (LO) \cite{5,6,7} and in 
NLO by us together with M. Klasen \cite{8} and very recently in \cite{9}. 
In the meantime the results of several analyses of H1 \cite{10,10b,11,12}
and ZEUS \cite{13} measurements of jet production in various ranges of
non-vanishing $Q^2$, namely $0.2 < Q^2 < 4~GeV^2$ \cite{13},
$1.4 < Q^2 < 25~GeV^2$ \cite{11}, $1.6<Q^2<80~GeV^2$ \cite{10b}
and $5 < Q^2 < 100~GeV^2$ \cite{12} have been published or presented
at workshops.

As long as the non-vanishing $Q^2 \ll \mu^2$, it is justified for calculating 
jet production to introduce a resolved contribution in addition to the direct
contribution in the same way as it has been done in the real photoproduction
case. Then the parton distributions of the photon depend on $x$ and
the scale $\mu^2 $, as in real photoproduction ($Q^2 = $0), and
in addition on the virtuality $Q^2$. Several models exist for describing
the $\mu^2$ evolution of these parton distribution functions (PDF's) with
changing $Q^2$ \cite{5, 14, 15}, but very little data from deep-inelastic
$e\g ^*$ scattering with photons $\g ^*$ of virtuality $Q^2 \neq 0$ 
\cite{16} exist, where they could be tested. Experimental data from
jet production in the region $Q^2 \ll \mu^2 \sim E_T^2$ could help
to gain information on the $Q^2$ evolution of these
photon structure functions. Parton densities of the virtual photon are
suppressed \cite{5, 14, 17} with increasing $Q^2$ and are, in the usual LO
definition, assumed to vanish like $\ln(\mu^2/Q^2)$ for
$Q^2 \rightarrow \mu^2$,  so that in the region $Q^2 \sim \mu^2$
the direct process dominates. Therefore, in the LO framework, it
depends very much on the choice of scale $\mu^2 $ in relation to $E_T^2$,
whether a resolved contribution is present in the region $Q^2 \geq E_T^2$.
This leads to a large scale dependence in addition to the usual scale
dependence of LO predictions and a treatment up to NLO is called
for. In analogy
to the photoproduction case in NLO the resolved and the direct contributions
are related to each other. So, if one adds for $Q^2 \neq 0$ a resolved
contribution, this has to be done in such a way that the contributions
involving the PDF's of the virtual photon 
are matched with the NLO direct photon contributions by
subtracting from the latter those terms which are already included
through the PDF's of the virtual photon. Such a subtraction has been
worked out in our earlier work with 
M. Klasen \cite{9} by separating the collinear photon initial state
singularities from the NLO corrections to the direct cross section.
There we studied inclusive one- and two-jet production with virtual
photons in the region $Q^2 \ll E_T^2$ by transforming to the HERA laboratory
system. In this system the results were compared to the photoproduction
cross sections and the unsubtracted direct-photon cross section up to NLO.
The dependence of the cross section on $Q^2$ had been investigated for
some cases up to $Q^2 = 9~GeV^2$. We found that with increasing $Q^2$ the
sum of the NLO resolved and the NLO direct cross section, in which the
terms already contained in the resolved part were subtracted,
approached the unsubtracted direct photon cross section. However, some
difference remained even at the highest studied $Q^2$. In NLO we expect
reduced scale dependence of the predictions, in particular, when also
the resoved cross section is calculated also up to NLO.

\section{NLO Calculation of Jet Cross Sections}

The NLO calculations are performed with the phase space slicing method.
As is well known the higher order (in $\alpha_s$) contributions to the
direct and resolved cross sections have infrared and collinear singularities.
To cancel them we use the familiar techniques. The singularities in the
virtual and real contributions are regularized by going to $d$ dimensions.
In the real contributions the singular regions are separated with the
phase-space slicing method based on invariant mass slicing. This way, we
have for both, the direct and the resolved cross section, a set of two-body
contributions and a set of three-body contributions. Each set is
completely finite, as all singularities have been canceled or absorbed
into PDF's. Each part depends separately on the phase-space slicing
parameter $y_s$. The analytic calculations are valid only for very small $y_s$,
since terms O($y_s$) have been neglected in the analytic integrations. For
very small $y_s$, the two separate pieces have no physical meaning. The $y_s$
is just a technical parameter which must be chosen sufficiently small and
serves the purpose to distinguish the phase space regions, where the
integrations are done analytically, from those, where they are performed
numerically. The final result must be independent of the parameter $y_s$. In
the real corrections for the direct cross section there are final state
singularities and contributions from parton initial state singularities
(from the proton side). This describes the calculation of the NLO cross
section for the full direct cross section as well as for the NLO resolved cross
section.

The resulting NLO corrections to the direct process become singular in the
limit $Q^2 \rightarrow 0$, i.e. direct production with real photons. For 
$Q^2 = 0$ these photon initial state singularities  are usually also
evaluated with the dimensional regularization method. Then the singular
contributions appear as poles in $\epsilon = (4-d)/2$ multiplied with the  
splitting function $P_{q\g }$ and have the form
$-\frac{1}{\epsilon} P_{q\g}$ multiplied with the LO matrix elements for
quark- parton scattering \cite{1}. These singular
contributions are absorbed into PDF's $f_{a/\g }(x)$ of the real photon.
For $Q^2 \neq 0$ the corresponding contributions are replaced by
\begin{equation}
 -\frac{1}{\epsilon} P_{q\g} \rightarrow -\ln(s/Q^2) P_{q\g}
\end{equation}
where $\sqrt{s}$ is the c.m. energy of the photon-parton subprocess. These
terms are finite as long as $Q^2 \neq 0$ and can be evaluated with $d=4$
dimensions. For small $Q^2$, these terms become large, which suggests to absorb
them as terms proportional to $\ln(M_{\g }^2/Q^2)$ in the PDF of the
virtual photon, which is present in the resolved cross section. $M_{\g }$ 
is the factorization scale of the virtual photon. By this absorption the
PDF of the virtual photon becomes dependent on $M_{\g }^2$, in
the same way as in the real photon case, but in addition it depends also on the
virtuality $Q^2$. Of course, this absorption of large terms is necessary only
for  $Q^2 \ll M_{\g }^2$. In all other cases the direct cross section
can be calculated without the subtraction and the additional resolved
contribution. $M_{\g }^2$ will be of the order of $E_T^2$. But also when
$Q^2 \simeq M_{\g }^2$, we can perform this subtraction. Then the
subtracted term will be added again in the resolved contribution, so that
the sum of the two cross sections remains unchanged. This way also the
dependence of the cross section on $M_{\g }^2$ must cancel, as long
as we restrict ourselves to the resolved contribution in LO only. 
The cross section with the subtractions in the NLO corrections to the direct
process will be denoted in the following the subtracted direct cross section.
It is clear that this cross section alone has no physical meaning. Only with
the resolved cross section added it can be compared with experimental data.

In the general formula for the deep-inelastic scattering cross
section, one has two contributions, the transverse ($d\sigma^U_{\g b}$)
and the longitudinal part ($d\sigma^L_{\g b}$). Since only the
transverse part has the initial-state collinear singularity we have performed
the subtraction only in the matrix element which contributes to 
$d\sigma^U_{\g b}$. Therefore we do not need the longitudinal PDF's 
$f^L_{a/\g }$. It is also well known that $d\sigma^L_{\g b}$
vanishes for $Q^2 \rightarrow 0$. The calculation of the resolved
cross section including NLO corrections proceeds as for real photoproduction
at $Q^2 = 0$ \cite{1}, except that the cross section is calculated also
for final state variables in the virtual photon-proton center-of-mass system.

The invariant mass resolution mentioned above is not suitable to distinguish
two and three jets in the final state. With the enforced small values for
$y_s$  the two-jet cross section would be negative in
NLO, i.e. unphysical. Therefore one chooses a jet definition that
enables one to define much broader jets. This is done in accordance with the
jet definition in the experimental analysis by choosing the jet definition
of the Snowmass meeting \cite{18}, where two partons are considered as
two separate jets or as a single jet depending on whether they lie
outside or inside a cone in the $(\eta ,\phi)$-plane with radius $R$ 
around the jet momentum. The cone parameter $R$ is chosen as in the
experimental analysis, in the following $R=1$. In NLO, the final state
consists of two or three jets.

\section{Parton Distributions of the Virtual Photon}

For the computation of the direct and resolved components in the one-
and two-jet cross sections we need the PDF's of the proton
$f_{b/P}(x)$ and of the photon $f_{a/\g }(x)$ 
at the respective factorization scales $M_P$ and $M_{\g }$. 
For the proton PDF's we have chosen
the CTEQ4M version \cite{19}.
The factorization scales are put equal to the renormalization 
scale $\mu$ ($M_{\g }=M_P=\mu$), where $\mu$ will be specified later. For
$f_{a/\g }$, the PDF of the virtual photon, we have chosen one of the
parametrizations of Schuler and Sj\"ostrand (SaS) \cite{15}. These sets are
given in parametrized form for all scales $M_{\g }$, so that they can be
applied without repeating the computation of the evolution. Unfortunately, 
these sets are given only in LO, i.e. the boundary conditions for
$Q^2=M_{\g }^2$ and the evolution equations are in LO. In  
\cite{14} PDF's for virtual photons have been constructed in LO and NLO.
However, parametrizations of the $M_\g$ evolution have not been worked
out. Second, these PDF's are only for $N_f = 3$ flavours, so that the charm
and bottom contributions must be added as an extra contribution.
Therefore we have selected a SaS version which includes charm and bottom as
massless flavours. We defined the subtraction of the collinear singularities
for the NLO direct cross section in the $\overline{\mbox{MS}}$ factorization. 
This has the consequence that, in addition to the dominant logarithmic term,
also terms (in the limit $Q^2 = 0)$ are left over in the NLO
corrections of the subtracted direct cross section (see \cite{8} for
further details). To be consistent we must use a parametrization
of the photon PDF that is defined in the $\overline{\mbox{MS}}$
factorization. In \cite{15} such PDF's in the $\overline{\mbox{MS}}$
scheme are given in addition to the PDF's in the DIS scheme,
where the finite parts are put equal to zero. Actually, this
distinction is relevant only in the NLO descriptions of the photon
structure function. Since numerically, however, it makes a
nonnegligible difference, whether one uses DIS or
$\overline{\mbox{MS}}$ type PDF's of the photon the authors of 
\cite{15} have  presented both types of PDF's. Unfortunately, the
$\overline{\mbox{MS}}$ version of \cite{15} is defined with the
so-called universal part of the finite terms, adopted from
\cite{20}. This does not correspond to the $\overline{\mbox{MS}}$
subtraction as we have used it in \cite{8}. Therefore we start with
the SaS1D parametrization in \cite{15}, which is of the DIS type with
no finite term in $F_2^{\g }(x,M_{\g }^2)$ and transform it with
the well-known formulas to the usual $\overline{\mbox{MS}}$ version.
The heavy quarks $c$ and $b$ are included as
massless flavours except for the starting scale $Q_0$, which is
$Q_0 = 600$ MeV for the $u$, $d$, $s$ quarks and the gluon and related to
the $c$ and $b$ quark masses, respectively.

\begin{figure}[bbb]
  \unitlength1mm
  \begin{picture}(122,80)
    \put(-4,-51){\epsfig{file=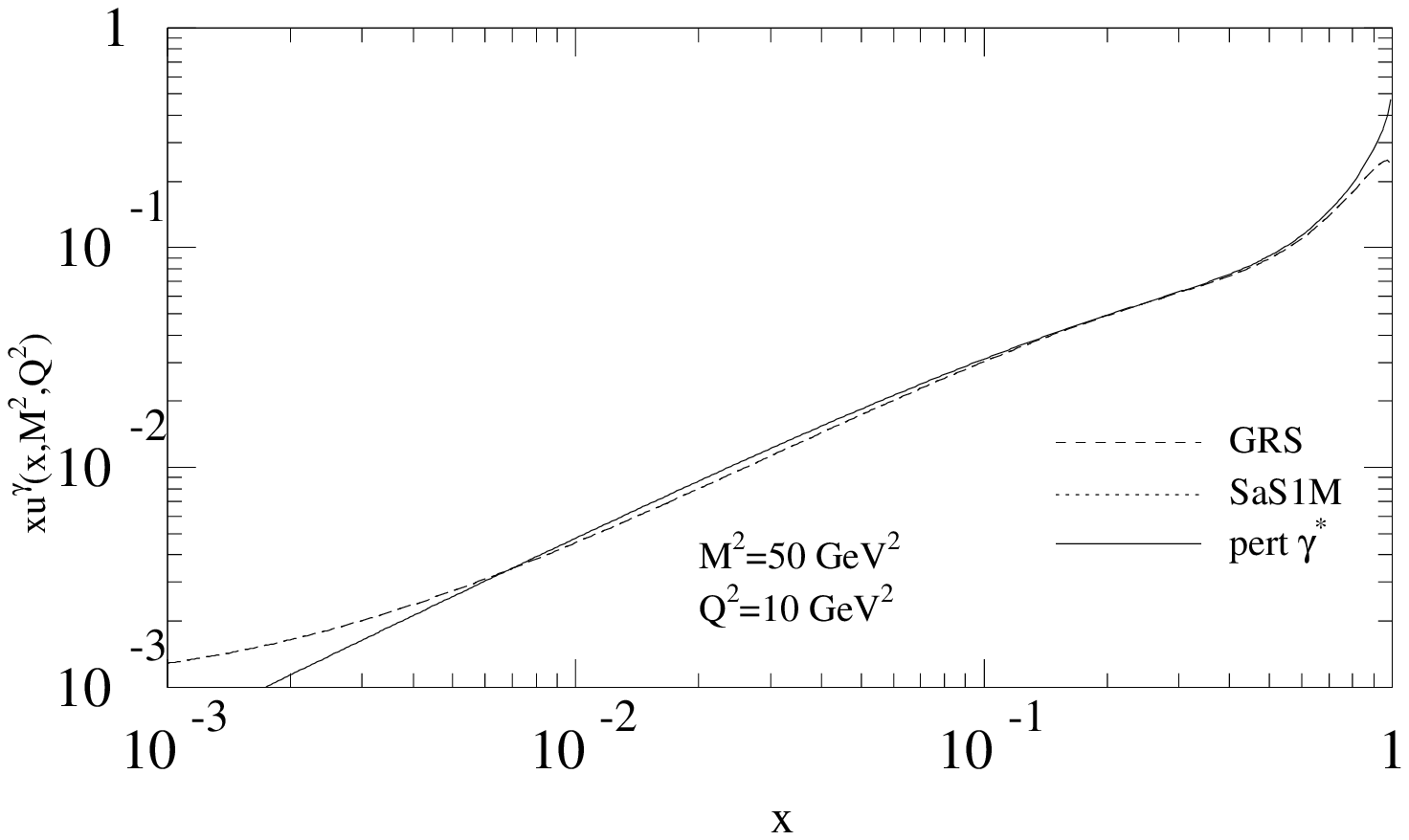,width=9.5cm,height=14cm}}
    \put(78,-51){\epsfig{file=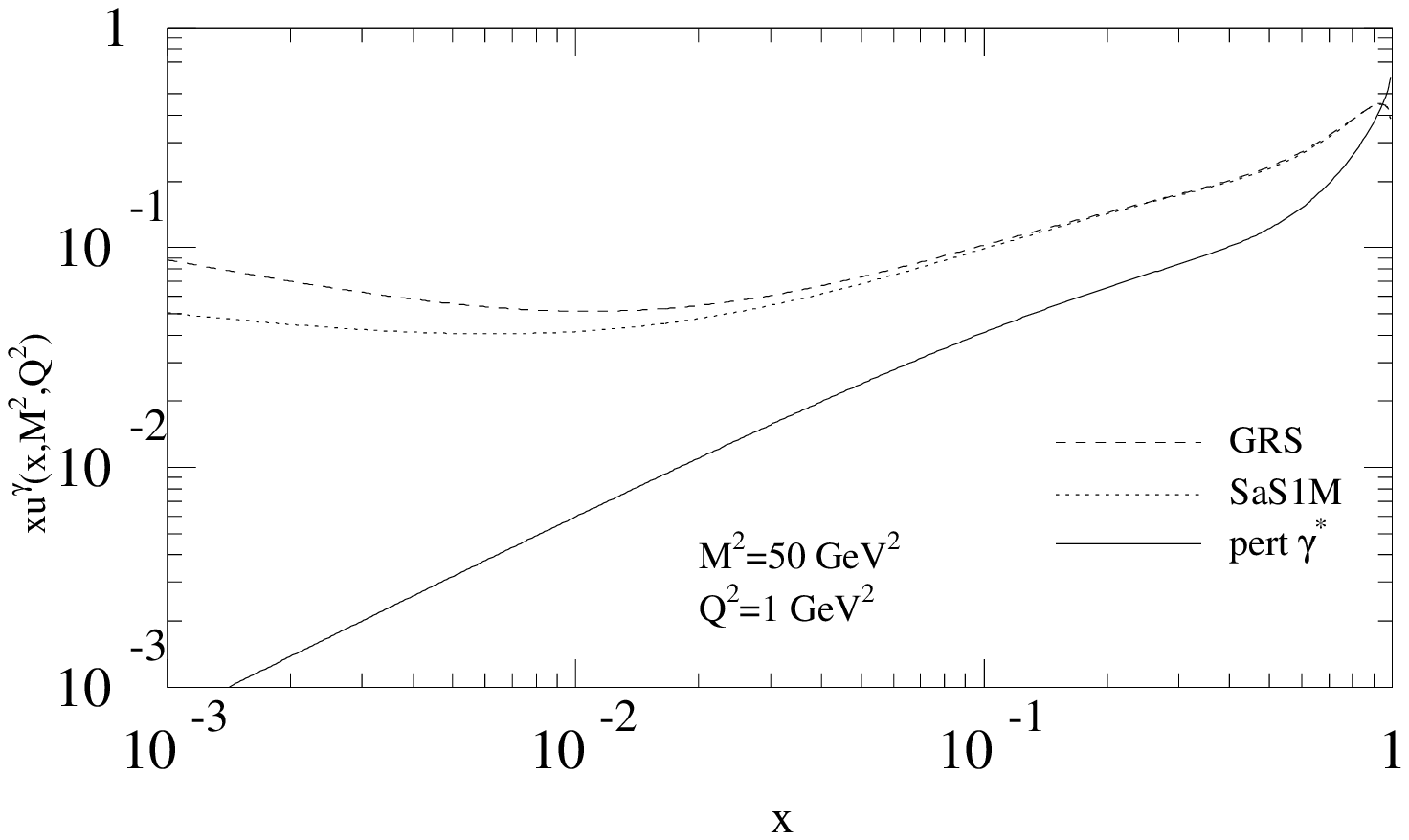,width=9.5cm,height=14cm}}
    \put(0,9){\parbox[t]{16cm}{\sloppy Figure 1: Comparison
        between the GRS and SaS1M LO predictions for the
        up-distribution of a virtual photon and the leading
        logarithmic singularity, denoted {\it pert $\g^*$},
        in the $\overline{\mbox{MS}}$ scheme at the scale
        $M^2=M_\g^2=50$ GeV$^2$ for $Q^2=10.0$ GeV$^2$ (left) and 
        $Q^2=1.0$ GeV$^2$ (right). }} 
  \end{picture}
\end{figure}
 
All the existing PDF's of the virtual photon are theoretical constructions,
which have not been tested or fitted to independent data from 
photon-photon scattering with a deep inelasic photon of large
virtuality which corresponds to the factorization scale $M_{\g}$ in
our application and a photon with a moderate virtuality (target
photon) which correspond to the $Q^2$ in our photoproduction case. In
order to do this we have to wait for data from LEP. The few points
measured by the PLUTO collaboration \cite{16} were reproduced by the
NLO PDF's of ref.~\cite{14}, which is unfortunately not available in
parametrized form. Actually the PLUTO measurement was only for
$Q^2=0.35~GeV^2$, which is not very much for testing the $Q^2$
dependence. In \cite{5,14,15} the PDF's are constructed in such a way
that they approach the PDF's for real photons in the limit $Q^2
\rightarrow 0$. The real photon PDF's have been compared to many data
for $F_2^{\g}$, so that the $x$-dependence
is constrained. The $Q^2$ dependence, however, is still untested and
depends on asumptions, in particular in connection with the hadronic part.
It is clear that for the larger $Q^2$ this is not relevant, since then the
PDF is given essentially by the pointlike or anomalous part. This is shown in
Fig. 1 (right), where we have plotted the $u$-quark part of the
virtual photon PDF for $Q^2=10~GeV^2$ and $M_{\g}^2 = 50~GeV^2$. Here
we compare the PDF of GRS \cite{5} with the SaS version \cite{15} and
with the leading logarithmic singularity approximation which is
identical to the subtraction term. As one can see these three
functions coincide except at very small $x$, where the PDF is very
small and in a small region around $x\simeq 1$. At smaller $Q^2$ the
situation is different. At $Q^2=1~GeV^2$, which is shown in Fig. 1
(left), the two PDF's, GRS and SaS are very similar for the
larger $x$, but differ very much from the logarithmic term.

\section{Results}

Here we shall show some numerical results in the form that
we present first the full direct cross section including the transversal
and the longitudinal part. Second, we have calculated the subtracted
direct cross section and the resolved cross section which we
superimpose to give the cross section which we compare with the full
direct cross section. These results are given for the inclusive one-jet cross
section, the exclusive two-jet rate and the inclusive two-jet cross
section as a function of the rapidity $\eta$. The exclusive two-jet
rate and the inclusive two-jet cross section will be compared with
recent H1 data \cite{12,11}. For the purpose of the comparison with
the exclusive two-jet data we have considered the $Q^2$ bins as shown
in Tab.\ 1. 
\begin{table}[bbb]
\renewcommand{\arraystretch}{1.6}
\caption{The seven subsequent bins of photon virtuality, $Q^2$,
  considered in this work.} 
\begin{center}
\begin{tabular}{|c|c|c|c|c|c|c|c|} \hline
 Bin number & I & II & III & IV & V & VI & VII \\ \hline 
 $Q^2$-range in GeV$^2$ & $[1,5]$ & $[5,11]$ & $[11,15]$ & $[15,20]$ &
 $[20,30]$  & $[30,50]$ & $[50,100]$ \\ \hline
\end{tabular}
\end{center}
\renewcommand{\arraystretch}{1}
\end{table}
In the experimental analysis only the bins II to VII are considered. We 
added the bin I in order to have results for cross sections of
rather small virtuality, where the resolved part is more important than for
all other bins. The bins chosen for the two-jet rate analysis of H1 involve
some further cuts on the scattering angle and the energy of the electron in
the final state. These are taken into account when we
compare with the experimental data. For the more theoretical comparisons
we have chosen simple cuts on the variable $y$, which is limited to the
region $0.05 < y < 0.6$.

Of some importance is the choice of the scale $\mu $. In bin I we have 
$Q^2 \ll E_T^2$, since in all considered cases $E_T>E_{T_{min}} > 5~GeV$, so
that $\mu = E_T$ would be a reasonable choice. Starting from bin V,
$Q^2 \geq E_{T_{min}}^2$, so that from this bin on with the choice $\mu = E_T$ 
the resolved cross section would disappear at the minimal $E_T$ and above
up to $E_T^2 = Q^2$. In order to have a smooth behaviour for all $E_T$ we have
chosen $\mu^2 = Q^2+E_T^2$, so that always $\mu^2/Q^2 > 1$ and in all bins
a resolved cross section is generated. Of course, in the sum of the
resolved and the subtracted direct cross section this scale
dependence, which originates from the factorization scale dependence at
the photon leg cancels to a very large extent in the summed cross
section. Only the NLO corrections to the resolved cross
section do not participate in the cancellation \cite{21,8}.
 
Some characteristic results for the one-jet cross section as a function of
$Q^2$ are shown in Fig 2 a, b, c and d. We have plotted the results for four
selected bins I, II, V and VII. In these figures we show the rapidity
distributions (the $E_T$ distributions are found in \cite{9}) 
\begin{equation}
  \frac{d\sigma^{1jet}}{d\eta} = \int dE_T
  \frac{d^2\sigma^{1jet}}{dE_Td\eta} \quad , 
\end{equation}
where we have integrated the
differential cross section over $E_T \geq 5~GeV$.

\begin{figure}[ttt]
  \unitlength1mm
  \begin{picture}(122,140)
    \put(-1,15){\epsfig{file=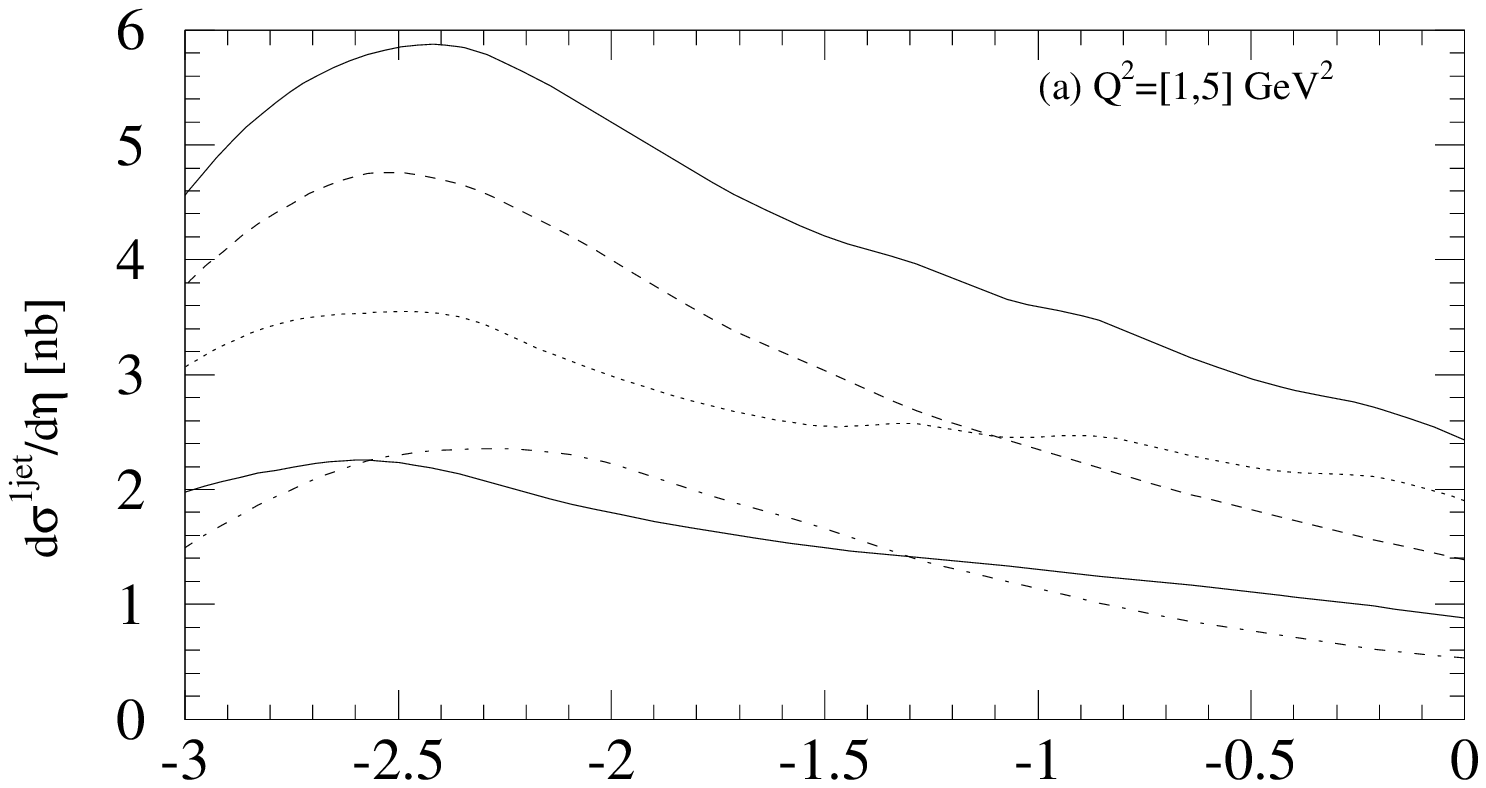,width=9cm,height=13.2cm}}
    \put(78,15){\epsfig{file=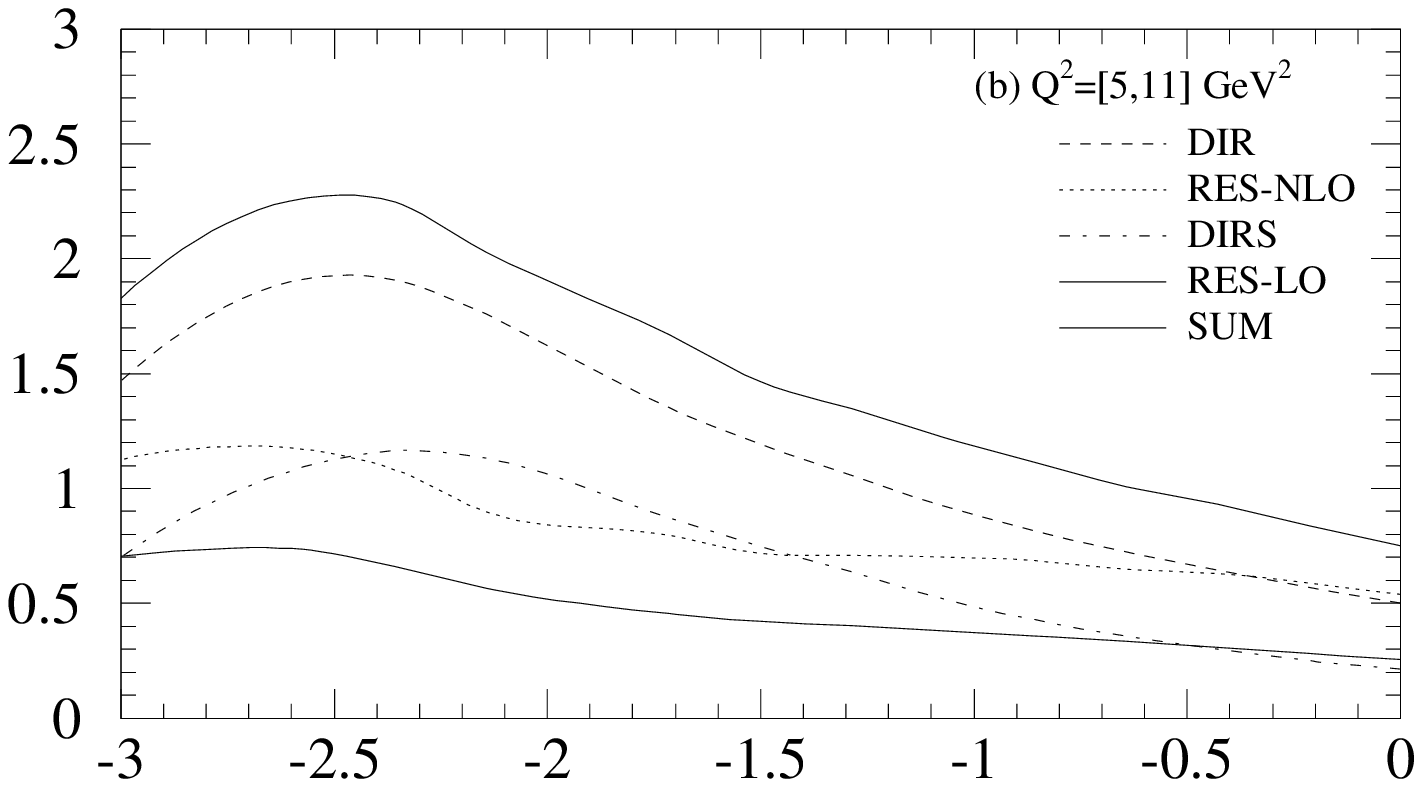,width=9cm,height=13.2cm}}
    \put(-1,-38){\epsfig{file=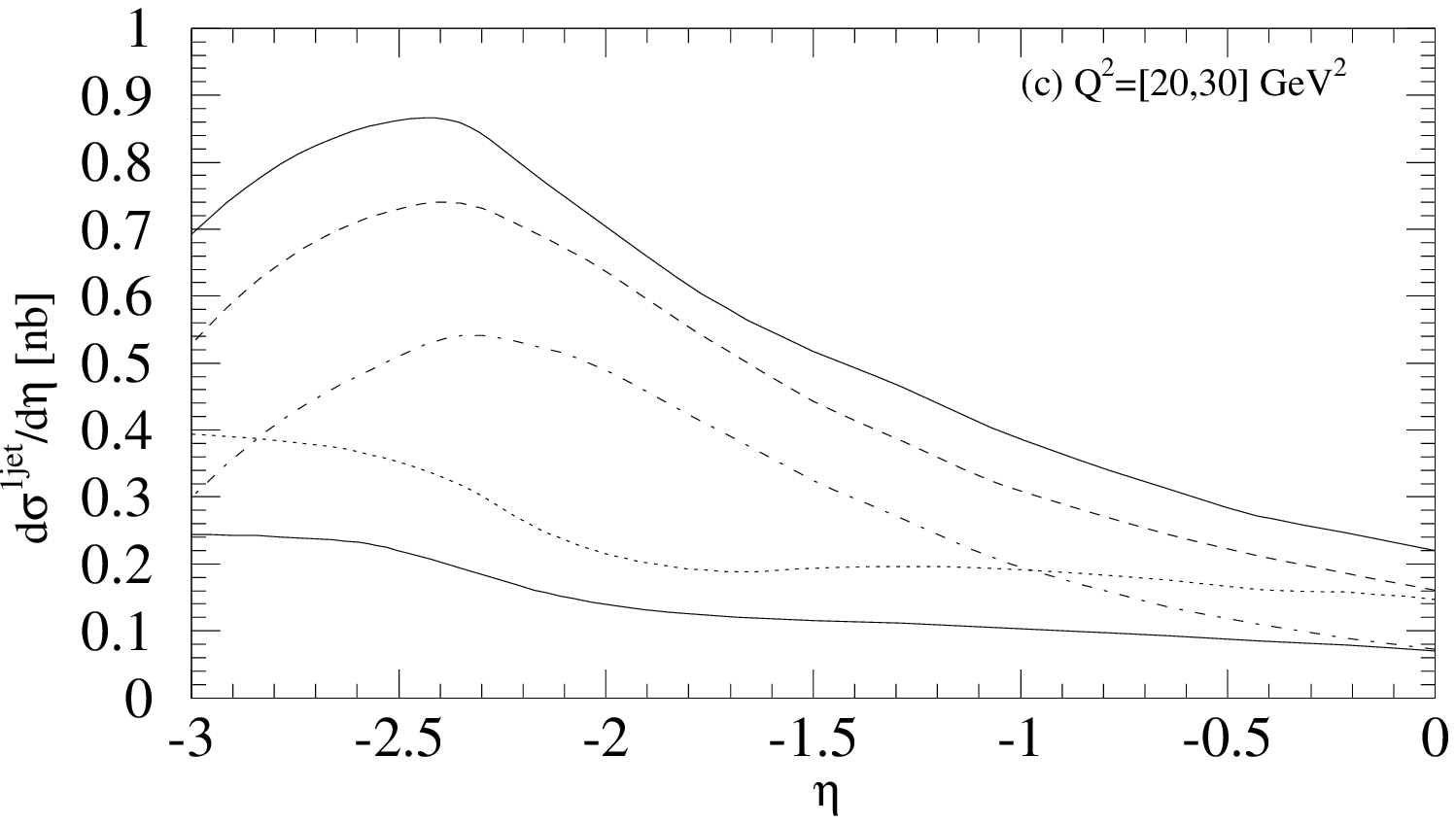,width=9cm,height=13.2cm}}
    \put(78,-38){\epsfig{file=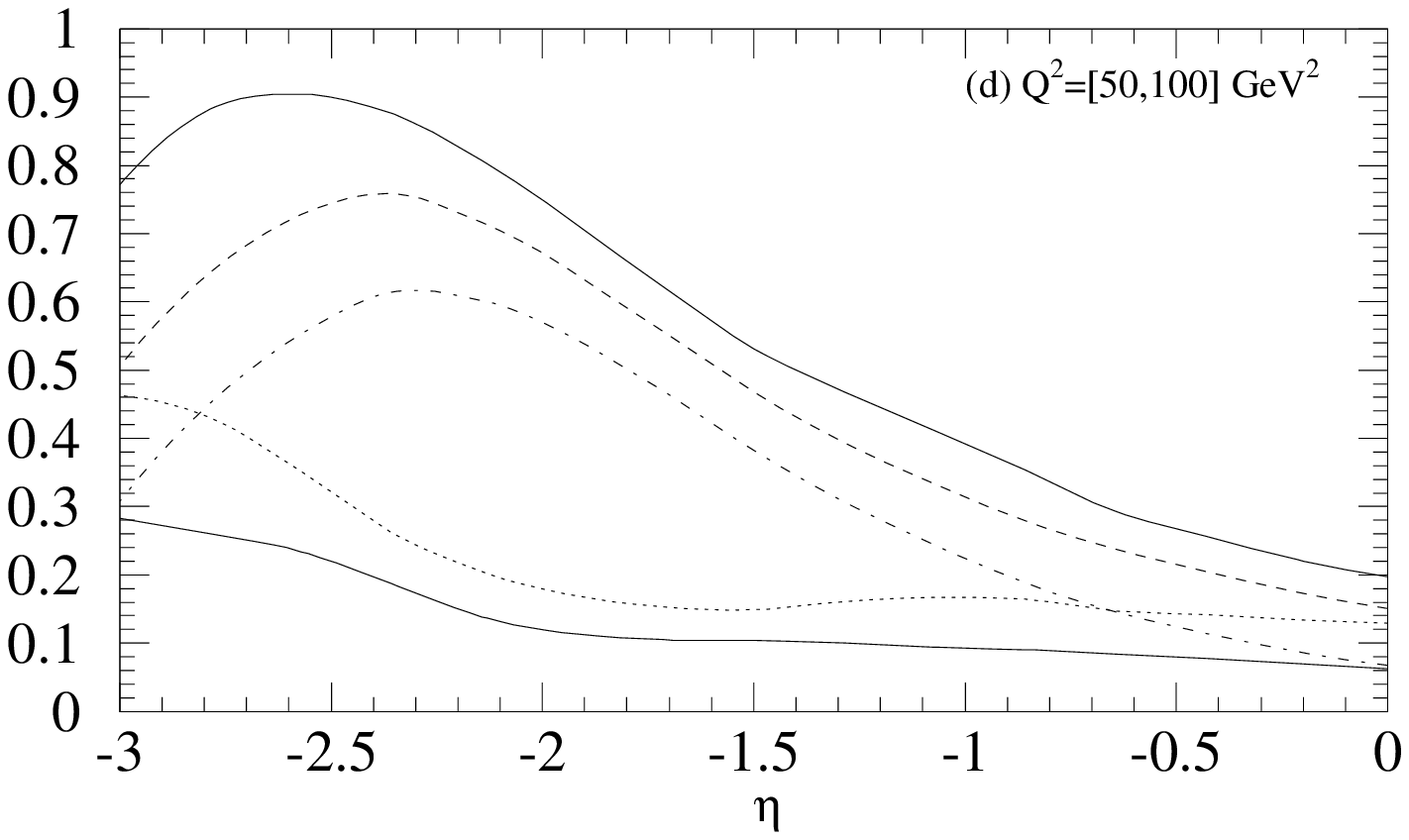,width=9cm,height=13.2cm}}
    \put(0,20){\parbox[t]{16cm}{\sloppy Figure 2: Inclusive
        single-jet cross section $d\sigma^{1jet}/d\eta$ integrated over
        $E_T>5$ GeV as a function of $\eta$. In (a): $1<Q^2<5$
        GeV$^2$; in (b): $5<Q^2<11$ GeV$^2$; in (c): $20<Q^2<30$
        GeV$^2$; in (d): $50<Q^2<100$ GeV$^2$. DIR stands for the NLO
        direct, DIRS is the NLO subtracted direct and RES-LO and
        RES-NLO are the LO and NLO resolved contributions, the lower
        full curve is always RES-LO, the upper one is SUM.}}
  \end{picture}
\end{figure}

In the four plots we show five curves for $d\sigma^{1jet}/d\eta$ as a function 
of $\eta $ in the range $-3 < \eta < 0$, since the cross section is
significantly large only in the backward direction $\eta < 0 $. $\eta
$ is the rapidity in the hadronic center-of-mass system. The five
curves present the resolved cross sections (denoted by RES) in LO and
NLO, the subtracted direct cross section denoted DIRS, the sum of DIRS and
the NLO resolved cross section, denoted SUM in the figures, and the
unsubtracted direct cross section labeled DIR. This cross section
should be compared to the cross section, labeled SUM (upper full curve). 
As we can see, for all four $Q^2$ bins the DIR cross
section in always smaller than the cross section obtained from the sum
of DIRS and the NLO resolved cross section. Near the maximum of the
cross sections they differ by approximately $25\%$ in bin I 
and by $20\%$ in the other bins. This means, at the respective $Q^2$ 
characterizing these bins, the summed cross section is always
larger than the pure direct cross section. This difference originates
essentially from the NLO corrections to the resolved cross section, as is
obvious when we add the LO resolved curve to the DIRS contribution in
Figs. 2 a, b, c and d. Up to a few percent the full DIR cross section and
the LO resolved plus subtracted direct cross section section are equal, except
at the lowest $Q^2$ bins.
This means that the term subtracted in the direct cross section is replaced
to a very large extent by the LO resolved cross section. Differences between
these two come from the evolution of the subtraction term to the scale $\mu
=\sqrt{Q^2+E_T^2}$ and contibutions of the hadronic part at low $Q^2$. 
This is to be expected since at the considered 
values of $Q^2>1$ GeV$^2$ the virtual photon PDF is 
essentially given by the anomalous (or point-like) part as shown in Fig. 1. 
All other contributions are of minor importance. Obviously the 
compensation of the LO resolved by the subtraction term is only
possible, if the photon PDF is chosen consistently with the
$\overline{\mbox{MS}}$ subtraction scheme, which is the case in our analysis.
This also explains that the inclusion
of the NLO corrections to the resolved cross section brings in
additional terms and that the sum of DIRS and the NLO resolved part
lies above the pure direct cross section. We conclude that except for
the lowest two $Q^2$ bins, the NLO direct cross section gives approximately 
the same results as SUM, if we restrict ourselves to the LO contributions of
the resolved cross section.
Similar results for the inclusive dijet cross section are shown in \cite{9}.
The differential cross section $d^3\sigma/dE_Td\eta_1d\eta_2$ yields the
maximum of information possible on the parton distributions and is better
suited to constrain them than with  measurements of inclusive single jets.

We now come to the comparison with the exclusive two-jet rate $R_2$ as
measured by H1 \cite{12}. $R_2$ measures the cross section for two-jet
production normalized to the total $ep$ scattering cross section in the
respective $Q^2$ bin. The data were obtained in the bins II to VII by
requiring for both jets $E_T   > 5~GeV$ in the hadronic center-of-mass frame 
with the additional constraints $y > 0.05,~k_0' > 11~GeV$ ($k_0'$ is the final
state electron energy), $156^{\circ}< \theta_e < 173^{\circ}$ and integrated
over $\eta_1$ and $\eta_2$ with $\Delta \eta=|\eta_1-\eta_2| <2$. 
Compared to the $Q^2$  bins considered in the previous sections, the H1 $Q^2$ 
bins are reduced through the additional constraints on $k_0'$ and the electron
scattering angle $\theta_e$. In the H1 analysis
the two jets are searched for with the usual cone algorithm with $R = 1$
applied to the hadronic final state. In addition $R_2$ measures the
exclusive two-jet rate, i.e. the contributions of more than two jets are not
counted (here we discard remnant jets). Symmetric cuts
$E_{T_1},E_{T_2} \geq 5~GeV$ are
problematic from the theoretical viewpoint since the so defined
cross section is infrared sensitive. With this same cut on the transverse
energy of both jets there remains no transverse energy of the third jet,
so that there is very little or no contribution from the three-body
processes. Through the phase space slicing, needed to cancel infrared
and collinear singularities in NLO, 3-body processes are always
included inside the cutoff $y_s$, which, however, are counted in the
$E_{T_1}=E_{T_2}$ contribution. For these contributions the $y_s$ cut 
acts as a physical cut. In order to avoid this sensitivity on $y_s$
one needs constraints on $E_{T_1},E_{T_2}$ or $E_{T_3}$ which avoids
the problematic region $E_{T_1}=E_{T_2}$. This problem was encountered
already two years ago in the calculation of the inclusive two-jet
cross section in photon-proton collisions \cite{22}. 

A possibility to remove the infrared sensitivity is to require
different lower limits on $E_{T_1}$ and $E_{T_2}$, as for example,
\begin{quote}
  $E_{T_1}, E_{T_2} > 5~GeV$, and
  if $E_{T_1}>E_{T_2}$ ($E_{T_2}>E_{T_1}$) then
  $E_{T_1}>7~GeV~ \\ (E_{T_2}>7~GeV)$,
\end{quote}
which we call the $\Delta$ mode. In this way, the third jet can have enough
transverse energy to avoid the infrared sensitivity. Other cuts that
also avoid the infrared region are possible but we will not discuss
them here (see \cite{9}). It is clear that the theoretical problems
with the $E_T$ cut on both jets appear equally in connection with NLO
corrections to the direct as well as to the resolved cross section. 
The $\Delta $ mode has been considered also recently in connection 
with inclusive two-jet photoproduction in the HERA system \cite{23}. 

Of course, the size of the dijet cross section depends on the way the
cuts on  $E_{T_1}$ and $E_{T_2}$ are introduced. Therefore, it is
important that the same cuts are applied in the theoretical
calculation and in the experimental analysis. The $\Delta$ mode, among
others, has been applied also in the measurements of $R_2$ \cite{12},
so that for this mode our results can be compared directly to the data. 
\begin{figure}[ttt]
  \unitlength1mm
  \begin{picture}(122,95)
    \put(-4,-36){\epsfig{file=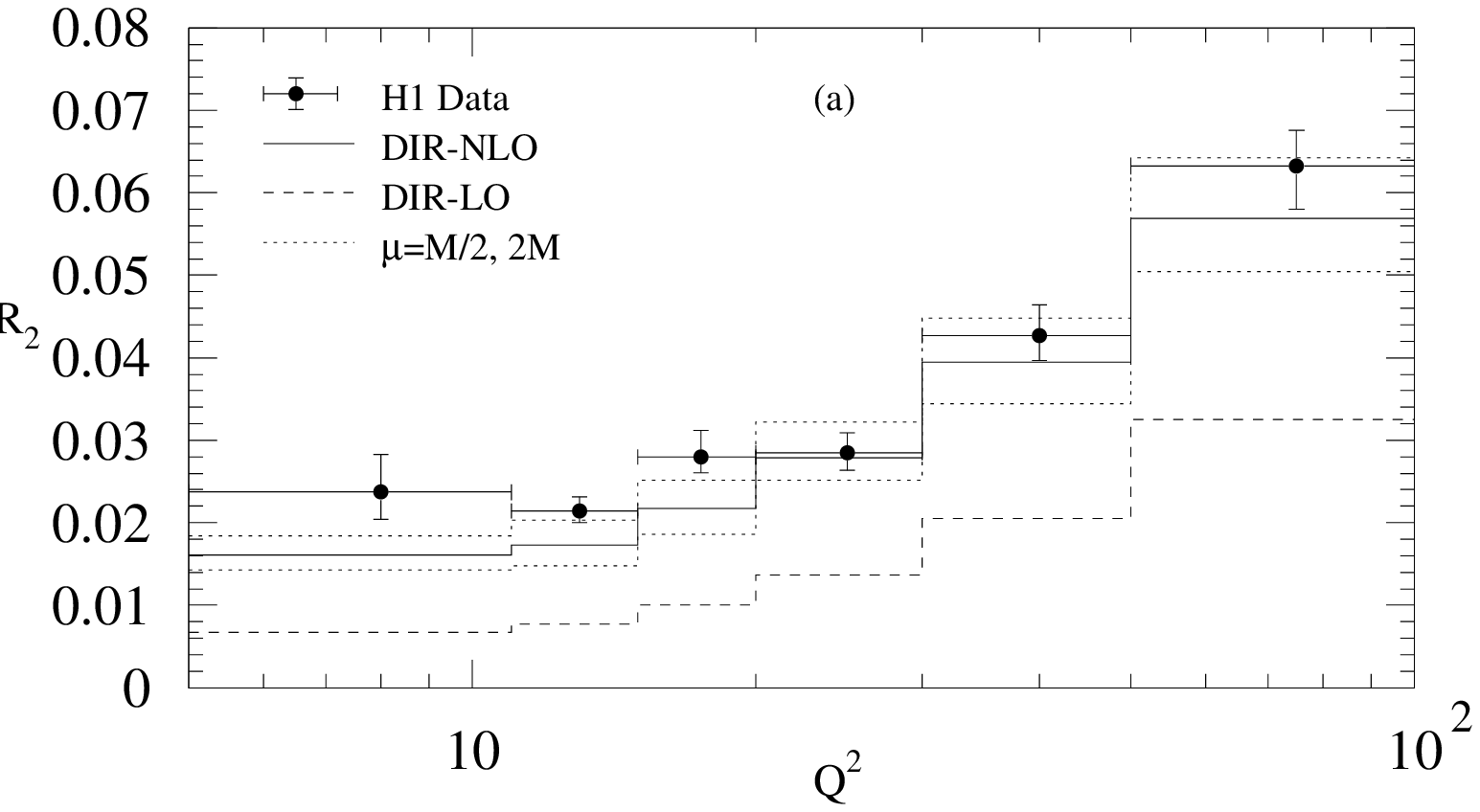,width=9.5cm,height=14cm}}
    \put(78,-36){\epsfig{file=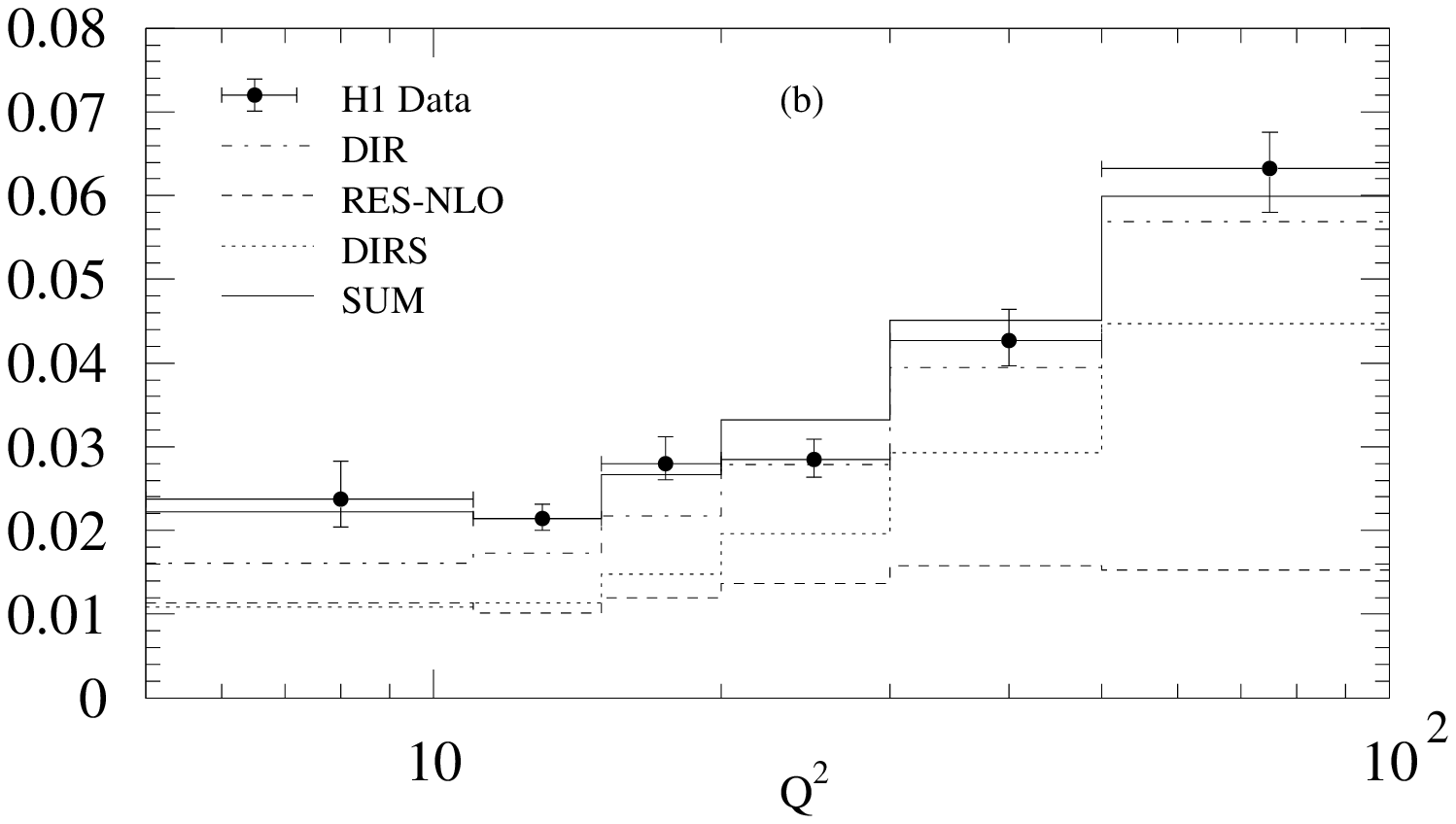,width=9.5cm,height=14cm}}
    \put(0,26){\parbox[t]{16cm}{\sloppy Figure 3: Dijet rate
        $R_2=\sigma^{2jet}/\sigma^{tot}$ with $E_{T_{min}}=5$ GeV 
        for the $\Delta$ mode compared to H1-data. (a) The full line
        corresponds to the NLO deep-inelastic dijet rate (DIR-NLO),
        the dashed curve gives the LO deep-inelastic dijet rate
        (DIR-LO). The dotted lines show the 
        scale variation for the NLO direct, where the upper dotted curve
        corresponds to the smaller scale. (b) The dash-dotted curve
        gives the NLO direct (DIR-NLO), the dashed is NLO resolved
        (RES-NLO), the dotted is NLO DIRS and the full is SUM.}}
  \end{picture}
\end{figure}
This we do in Fig.\ 3. We compare results for
the direct cross section in LO (DIR-LO) and NLO (DIR-NLO) for three
different scales $\mu = M/2,~M,~2M$ where $M=\sqrt{Q^2+E_{T_1}^2}$,
calculated for the $Q^2$ 
bins II to VII with the additional cuts on $k_0'$ and $\theta_e$
mentioned above. We see that the NLO corrections are
appreciable. Since the scale $\mu $ is rather low we have to expect
such large K factors. On the other hand the scale variation is
moderate, so that we are inclined to consider the NLO cross section as
a safe prediction. In Fig.\ 3 b we compare the NLO
direct cross section (DIR) with the sum (SUM) of the subtracted direct (DIRS)
and the NLO resolved cross section (RES-NLO) for the six $Q^2$ bins. In
addition, we show the contribution of the two components (DIRS and
RES-NLO) in the sum separately. In the first $Q^2$ bin, DIRS and the NLO
resolved cross section are almost equal, the cross section in the
largest $Q^2$ bin is dominated by DIRS. In this bin the unsubtracted
direct cross section DIR is almost equal to the 
sum of DIRS and NLO resolved. In the first $Q^2$ bin this cross section is
$50\%$ larger than the NLO direct cross section. We also compare with the H1
data \cite{12}. In the smaller $Q^2$ bins the sum of DIRS and NLO resolved
agrees better with the experimental data than the DIR cross
section. In the two largest $Q^2$ bins the difference of the cross
sections DIR and SUM is small and it can not be decided which of these
cross sections agrees better with the data due to the experimental
errors. This is in contrast to the 30\% difference between the DIR and
SUM found for the inclusive single-jet cross sections in Fig. 2, which we
attribute to the NLO corrections of the resolved contributions. This
difference is reduced in the dijet rate $R_2$ due to the specific cuts
on the transverse energies of the two jets in the definition of the
dijet rate. These cuts suppress the resolved component stronger than
the direct one, which leads to the observed behaviour of the dijet
rate at the large $Q^2$ bins. We emphasize that the theoretical cross
sections are calculated on parton level whereas the experimental
two-jet rate is based on hadron jets. Corrections due to
hadronization effects are estimated to be typically around 5\% and at
most 20\% \cite{12}. We finally mention that we have also compared NLO 
predictions with a different mode that avoids the infrared region and
found  similar results as those for the $\Delta$ mode (see
\cite{9} for details).
\begin{figure}[ttt]
  \unitlength1mm
  \begin{picture}(122,138)
    \put(35,93){\epsfig{file=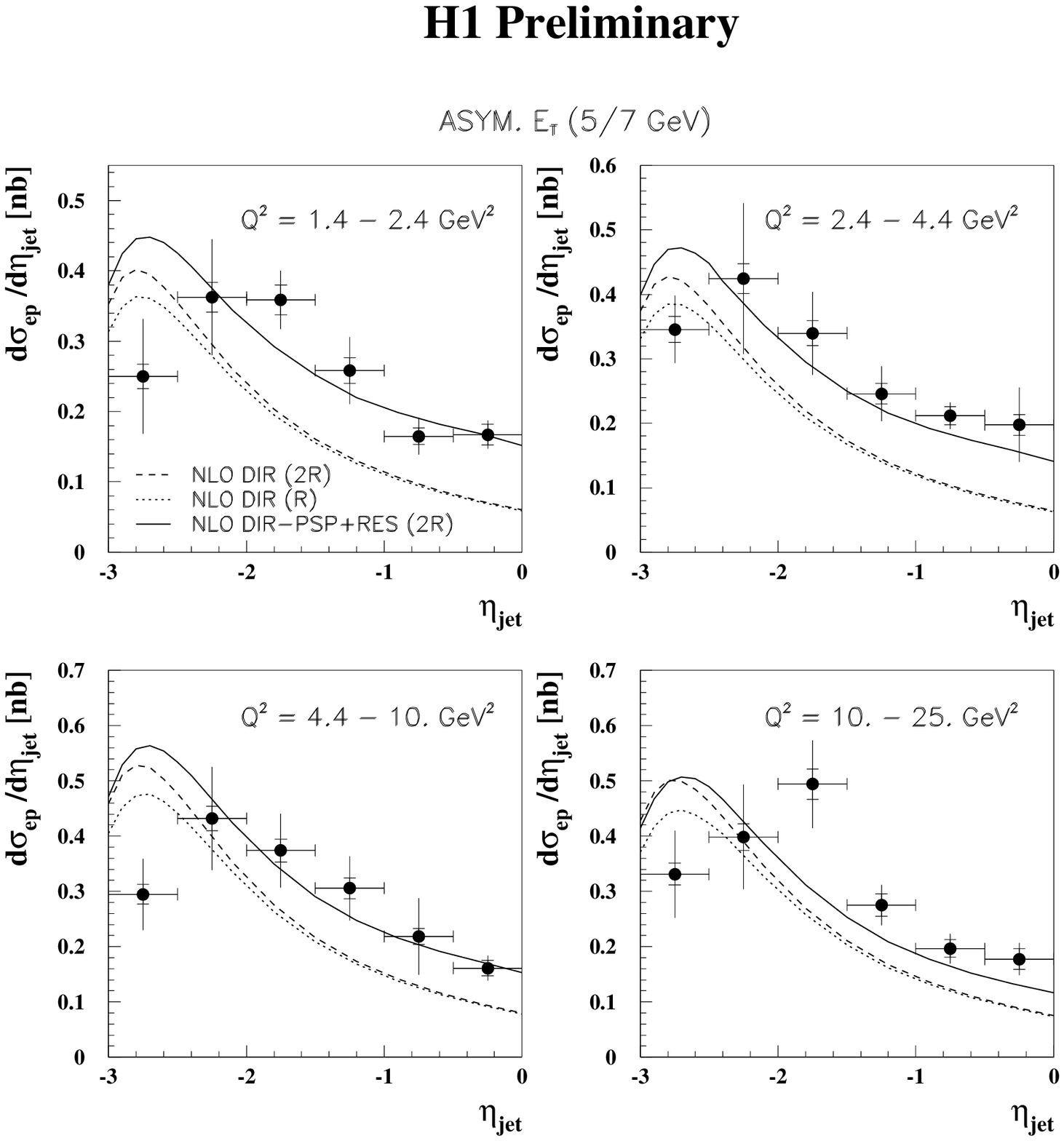,width=6.2cm,height=5.1cm,
        bbllx=42pt,bblly=514pt,bburx=347,bbury=746pt,angle=0,clip=}} 
    \put(0,8){\parbox[t]{16cm}{\sloppy Figure 4: Dijet inclusive 
        cross section $d\sigma_{ep}/d\eta_{jet}$ as a function of
        rapidity in the $\Delta$ mode compared to NLO direct (DIR) for
        two different $R_{sep}$ parameters and the sum of NLO DIRS plus
        NLO resolved (calculated with JETVIP \cite{9}). }}
  \end{picture}
\end{figure}

We come to a second analysis for dijet cross sections, presented in 
\cite{11}, which was also done in the hadronic center-of-mass
frame. Jets where selected using the cone algorithm with $R=1$. As
before, the $\Delta$ mode was used to avoid infrared sensitive
regions. Some additional cuts where opposed on the scattered electron
(see \cite{11}). The differential inclusive two-jet cross section
$d\sigma/d\eta$ as a function of the rapidity  for the
four $Q^2$ bins $1.4<Q^2<2.4~GeV^2$, $2.4<Q^2<4.4~GeV^2$, 
$4.4<Q^2<10.~GeV^2$ and $10.<Q^2<25.~GeV^2$ is shown in Figure 4. 
The full curve is the cross section obtained from the NLO resolved
plus subtracted direct cross section. The two other curves (dashed and
dotted) present two unsubtracted direct cross sections with two
different choices of the $R_{sep}$  parameter: $R_{sep}=R$ and
$R_{sep}=2\,R$. The latter choice correponds to no $R_{sep}$
parameter, which we had chosen also for all results presented so far. 
The influence of the $R_{sep}$ parameter is small except at very
negative $\eta $'s. The agreement between the data \cite{11} and the
NLO predictions is better when a resolved component is included,
especially in the forward $\eta$ region. We note that the agreement is
even better, in particular for $\eta \simeq -3$, when corrrections
for hadronisation are applied \cite{11}.

\section{Conclusions}

We have reviewed the calculations of cross sections in NLO for inclusive
single-jet and dijet production in low $Q^2$ $ep$ scattering at HERA.
The results of two approaches were compared as a function of $Q^2$ in
the range $1<Q^2<100$ GeV$^2$. In the first approach the jet
production was calculated in NLO from the usual mechanism where the
photon couples directly to quarks. In the second approach the
logarithmic dependence on $Q^2$ of the NLO corrections is absorbed
into the parton distribution function of the virtual photon and the
jet cross sections are calculated from the subtracted direct and the
NLO resolved contributions. Over the whole $Q^2$ range considered in this work,
this sum gives on average $25$\% larger single-jet cross sections than the
usual evaluation based only on the direct photon coupling. This
difference is attributed to the NLO corrections of the resolved cross
sections. If these NLO corrections are neglected the sum of the
subtracted direct and the LO resolved contributions agrees 
approximately with the unsubtracted direct cross sections. 

We calculated also the dijet rate based on the exclusive dijet cross
section and differential dijet distributions and compared it with recent
H1 data. The dijet rate is
plotted as a function of $Q^2$, the rapidities and transverse energies
are integrated with $E_T\ge 5$ GeV. The dijet rate is sensitive to the
way the transverse energies of the two jets are cut. If the cuts on the
$E_T$'s are exactly at the same value the cross section is infrared
sensitive. We showed results with one particular definition for
the kinematical constraints on the transverse energies of the measured jets.
The calculated and the measured two-jet rates agree quite well over the
measured $Q^2$ range $5<Q^2<100$ GeV$^2$. In the lowest $Q^2$ bin only the
dijet rate based on the sum of the subtracted direct and resolved
cross sections agrees with the experimental value. Also the differential dijet
cross sections as a function of rapidity agreed quite well with the
predictions. The version of the theory including a resolved component was
preferred by the data.

\subsection*{Acknowledgements}

We thank M.~Tasevsky for producing the NLO curves for Fig.~4 with 
JETVIP and for providing us with the plots including the H1 data. 
One of us (B.P.) thanks the organizers of the workshop for a pleasant
and stimulating working athmosphere.


\end{document}